\documentclass[prl,twocolumn,showpacs,preprintnumbers,amsmath,amssymb]{revtex4-1}
\usepackage{graphicx}

\def\lsim{\mathrel{\rlap{\lower4pt\hbox{\hskip1pt$\sim$}}
    \raise1pt\hbox{$<$}}}                
\def\gsim{\mathrel{\rlap{\lower4pt\hbox{\hskip1pt$\sim$}}
    \raise1pt\hbox{$>$}}}                

\begin{document}
\title{Scaling of random walk betweenness in networks}
\author{Onuttom Narayan$^1$ and Iraj Saniee$^2$}
\affiliation{$^1$ Department of Physics, University of California, Santa Cruz, CA 95064}
\affiliation{$^2$ Mathematics of Networks Department, Bell Laboratories, 
Alcatel-Lucent, 600 Mountain Avenue, Murray Hill, NJ 07974}
\date{\today}
\begin{abstract}
The betweenness centrality of graphs using random walk paths instead
of geodesics is studied. A scaling collapse with no adjustable parameters
is obtained as the graph size $N$ is varied; the scaling curve
depends on the graph model. A normalized random
betweenness, that counts each walk passing through a node only once,
is also defined. It is argued to be more useful and seen to have
simpler scaling behavior. In particular, the probability for a
random walk on a preferential attachment graph to pass through the root node is
found to tend to unity as $N\rightarrow\infty.$
\end{abstract}
\maketitle
\section{Introduction}
\label{sec:intro}

There is a large literature on the characterization of the minimal 
capacity required to meet end-to-end flows in networks.  This characterization
is typically expressed in terms of necessary or sufficient
conditions on link capacities, see~\cite{ff,efs}
for the single commodity and~\cite{lo,sm,pa,os} for multi-commodity flows. 
Alternatively, one can consider the case when the flows are as efficient as 
possible, i.e. along shortest paths, and capacities have to be chosen to 
accommodate them.
Depending on the structure of a network, the number of
such shortest paths that pass through a node can differ substantially
from node to node. As an obvious example, if two halves of
a network are linked by a narrow `bridge', a large number of shortest
paths will pass through the bridge. A more subtle example is that
of `hyperlattices' or hyperbolic networks~\cite{rno}, where
despite the absence of such a bridge, shortest paths coalesce
disproportionately near the center or core of the
network~\cite{oniis,jonck}. In social networks, if one considers
the shortest paths between all source-destination pairs of nodes
in the network, the number of paths passing through a node is referred to
as the ``betweenness centrality'' or betweenness of the node~\cite{src}, and is a
global measure of the connectivity at the node. For communications
networks, this quantity usually measures the load or congestion at the node
due to unit end-to-end flows. Although the
physical significance of the two is different, they are the same except when 
there are multiple shortest paths between node pairs~\cite{brandes}.

There are natural settings in which one might deviate from shortest
path routing. Longer paths may be used for load balancing by going
around highly loaded regions. In the case of capacitated networks,
departures from shortest path routing can avoid network expansion.
In order to study the behavior of networks when shortest path routing
is no longer used, it is useful to consider its opposite extreme,
random walk routing. 

To be specific, we consider a graph with $N$ nodes, with the same rate
of traffic flow between all $N(N-1)$ possible source-destination pairs. 
The dynamics are discrete time. At every time step, one packet of traffic is
injected at each node for each other node as destination. Any packet of 
traffic that was already in the graph moves randomly with equal probability
to one of the adjoining nodes. If a packet reaches its destination, it is
removed from the network at the next time step. The network is assumed to 
consist of one connected component.

Random walks on networks have been studied earlier using the graph
Laplacian~\cite{dorog,rieger} with methods similar to those we will
use in this paper, but the quantities studied are different. In
particular, Ref.~\cite{rieger} defines a random walk centrality by
computing the average time to travel from a source to a destination
node, and the change when the source and destination are reversed.
We use alternative definitions in this paper. Ref.~\cite{dorog}
investigates the spectral gap of the graph Laplacian.

With random walk routing, it is no longer definite whether a path
from a source to a destination will pass through some node; each
node has a probability of being on the path. Also, the load and the
betweenness centrality are no longer equivalent.  This is because
a random walk wandering through the network can pass through a node
several times.  While it is appropriate to count each of these
traversals as contributing one unit to the load at the node, it is
unreasonable to consider them as each adding to the betweenness of
the node.

In a separate paper~\cite{diff1}, we have shown that the load at
each node with random walk routing is linearly dependent on the
degree of the node, with a proportionality constant $N \sum_\alpha
1/\lambda_\alpha$ involving the sum of the inverses of the non-zero
eigenvalues of the Laplacian on the graph. We computed how the
proportionality constant scales with $N$ for various network models.
In this paper, we consider different ways to define the random walk 
betweenness of the nodes and the scaling thereof for various network 
models.

In the next section of this paper, we consider the definition of the random 
walk betweenness due to 
Newman~\cite{newman}. We obtain an
expression for this quantity in terms of the eigenfunctions of the
Laplacian, from which we numerically obtain the distribution of
random betweenness as a function of the number of nodes $N$ in the
network.  By mapping the random walk problem to current flowing in
an electrical circuit, a prescription to achieve a scaling collapse
of the distribution is obtained and verified for simple lattice
graphs such as square and triangular lattices.  This relies on the 
fact that the continuum limit for current flow on these graphs is 
diffusion on a plane. Surprisingly, the
same prescription works for Erdos Renyi~\cite{er} and extended
Barabasi Albert~\cite{barabasi} graphs even though there is no 
underlying continuum limit. Thus for all the graph models,
a scaling collapse is obtained with no adjustable parameters, using only
the measured average distance between node pairs $l_N$ in a $N$-node 
graph.  The scaling form fails for hyperbolic grids ---
discretizations of the Poincare disk --- implying that it is not
trivially true.


In Section~\ref{sec:alternate} we present an alternative definition
of the random walk betweenness that we argue is more appropriate,
in that a random walk contributes only once to the betweenness of
every node it passes through, regardless of the number of times it
does so.  With this definition, an even simpler scaling collapse
is obtained for the betweenness distribution as a function of the 
graph size $N.$

\section{Random walk betweenness}
\label{sec:rwbetw}
As in Ref.~\cite{newman}, we consider the random walk process
described in the Introduction: with discrete time dynamics, one
random walker is injected into the network at time $t=1,2,3\ldots$
at each source node for each other destination node; thus there are
$(N-1)$ walkers injected at each node at each time step. Any walker
that is present at node $i$ at time $t$ is removed from the network
at time $t+1$ if $i$ is its destination. If not, one of the neighbors
of $i$ is chosen randomly, and the walker moves there at time $t+1.$
The probability of choosing each of the neighbors of $i$ is $1/d_i,$
where $d_i$ is the degree of the $i$'th node. Note that a walker
that returns to its source as it moves around randomly continues
as it would from any other node.

For any given source node $k$ and destination node $l,$ the {\it
net\/} time-averaged current through each of the edges connected
to a node $i$ is computed, and the magnitudes of all these are
added.  After adding the magnitudes of the net currents for the
edges connecting to node $i,$ the result is averaged over all source
nodes $k$ and destination nodes $l$ to define the random walk
betweenness of $i.$ With this definition, if a random walk that
reaches node $i$ moves out at the next time step to node $j$ through
the edge $(ij),$ and returns to $i$ through the edge $(ji)$ at a
later time, the outward flow along $(ij)$ and the return along
$(ji)$ cancel each other.  However, if the random walk leaves the
node $i$ along the edge $(ij)$ and returns later through a different
edge $(ki),$ the two do not cancel out, but are instead added.
Although this definition only partially cancels the effect of a
random walk looping through a node multiple times, it has the
advantage that it can be mapped to an electrical circuit and solved
using Kirchoff's laws~\cite{newman}.

To obtain an analytical expression for the random walk betweenness,
we follow the approach of Ref.~\cite{diff1}. For the source $k$ and
destination $l,$ let $\tilde p_i^{kl}(t)$ be the number of walkers
at node $i$ at time $t,$ averaged over all the random paths that
the walkers can take. Let $A_{ij}$ be the adjacency matrix of the
graph. Then
\begin{equation}
\tilde p_i^{kl}(t + 1) = 
\delta_{ik} + \sum_{j\neq l} A_{ij} \frac{\tilde p_j^{kl}(t)}{d_j}.
\label{eq1}
\end{equation}
The first term on the right hand side accounts for the fact that
one walker is injected at node $k$ for destination $l$ at each time
step. The second term represents the walkers that move to node $i$
at time $t+1$ from adjacent nodes at time $t.$ The sum in this term
excludes the node $l$ because any walker that was at the node $l$
(the destination) at time $t$ is removed from the network and is
no longer present at time $t+1.$

We define $p^{kl}_i = (1 - \delta_{il}) \tilde p^{kl}_i.$ In other
words, $p^{kl}_i = \tilde p^{kl}_i$ except for the destination node,
$i=l,$ where $p^{kl}_l = 0.$ The sum in Eq.(\ref{eq1}) can now be
unrestricted for $i\neq l.$ The rate equation for the $p_i$'s is
\begin{equation}
p^{kl}_i(t + 1) = \delta_{ik} + \sum_j A_{ij} \frac{p^{kl}_j(t)}{d_j}
\label{ee2}
\end{equation}
for $i\neq l,$ with the boundary condition $p^{kl}_l(t+1) = 0.$ 

In steady state, we know that the load flowing into the node $l$
at any time step must be equal to the load injected into the node
$k,$ i.e. unity. Therefore $\sum A_{lj} p^{kl}_j = 1,$ and we can
extend Eq.(\ref{ee2}) in steady state as
\begin{equation}
p^{kl}_i= \delta_{ik} - \delta_{il} + \sum_j A_{ij} \frac{p^{kl}_j}{d_j}
\label{ee3}
\end{equation}
for all $i.$ This is a degenerate set of equations because the
matrix $A_{ij} - d_i\delta_{ij}$ has zero determinant; if $p^{kl}_i$
is a solution to the equation, so is $p^{kl}_i+ a d_i$ for any $a.$
Therefore, even though the node $l$ is in the domain of validity
of Eq.(\ref{ee3}) unlike (\ref{ee2}), we can still impose the 
condition $p^{kl}_l= 0.$

In order to convert Eq.(\ref{ee3}) to a Hermitean eigenvalue problem,
we define $p^{kl}_i = d_i r^{kl}_i$ and $L_{ij} = d_i\delta_{ij} -
A_{ij}.$ Then
\begin{equation}
\sum_j L_{ij} r^{kl}_j = \delta_{ik} - \delta_{il}
\label{ee4}
\end{equation}
with $r^{kl}_l = 0.$ Here $L_{ij}$ is the Laplacian for the graph.
Since $L_{ij}$ is a real symmetric matrix, it has a complete set
of real eigenvalues $\lambda_\alpha$ and real orthonormal eigenvectors
$\xi^\alpha$ for $\alpha = 0, 1, 2\ldots N - 1.$ Using the standard
properties of the graph Laplacian, all the eigenvalues are non-negative,
and since the graph has been assumed to have one component, there
is only one zero eigenvalue $\lambda_0$ with normalized eigenvector
$\xi^0 = (1, 1, 1,\ldots 1)/\sqrt N.$ Projecting both sides of
Eq.(\ref{ee4}) onto each eigenvector $\xi^\alpha$ (note that the
projection of the right hand side onto $\xi^0$ is zero), one can
verify that
\begin{equation}
r^{kl}_j = \sum_{\alpha = 1}^{N-1} \frac{\xi^\alpha_k 
- \xi^\alpha_l}{\lambda_\alpha}\xi^\alpha_j + c^{kl} \xi^0_j,
\end{equation}
where $c^{kl}$ remains to be determined from the condition $r^{kl}_l
= 0.$ Since $\xi_j^0$ is independent of $j,$ we have
\begin{equation}
r^{kl}_j = \sum_{\alpha = 1}^{N-1} \frac{
\xi^\alpha_k - \xi^\alpha_l}{\lambda_\alpha}\xi^\alpha_j  
- \sum_{\alpha = 1}^{N-1} \frac{1}{\lambda_\alpha}
[\xi^\alpha_k\xi^\alpha_l - (\xi^\alpha_l)^2].
\label{qj}
\end{equation}

The net inflow to any node $i$ from a neighbor $j$ is then $A_{ij}
p_j^{kl}/d_j - A_{ji} p_i^{kl}/d_i = A_{ij} (r_j^{kl} - r_i^{kl}).$
Adding the magnitudes of the currents along all the edges attached
to the node $i$ (with an extra factor of half), 
\begin{equation}
b_i^{kl} = {1\over 2}\sum_j A_{ij} |r_i^{kl} - r_j^{kl}| 
+ {1\over 2} (\delta_{ik} + \delta_{il})
\label{ieq}
\end{equation}
where the second term on the right hand side is due to the current
flowing into and out of the graph at the nodes $k$ and $l$ respectively.
The random walk betweenness of the node $i$ is then obtained by
averaging over all possible sources and destinations
\begin{equation}
b_i = \frac{1}{2 N(N-1)} \sum_{k\neq l} 
\sum_j A_{ij} |r_i^{kl} - r_j^{kl}| + \frac{1}{N}.
\label{beq}
\end{equation}
Using Eq.(\ref{qj}), this yields
\begin{equation}
b_i = \frac{1}{2 N (N - 1)} \sum_{k\neq l} \sum_j 
A_{ij}\sum_{\alpha\neq 0}\Bigg\vert (\xi^\alpha_k - \xi^\alpha_l) 
\frac{1}{\lambda_\alpha} (\xi^\alpha_i - \xi^\alpha_j)
\Bigg\vert + \frac{1}{N}.
\label{betw1}
\end{equation}

The sum over $\alpha$ on the right hand side would be equivalent
to a linear combination of matrix elements of $L^{-1}$ if the matrix
were invertible, but it is not. To circumvent this problem, we
define the operator $M = L + P,$ where $P$ is the projection operator
onto the zero eigenvector of the Laplacian: $P_{ij} = 1/N.$ Unlike
the Laplacian, $M$ is invertible, and
\begin{equation}
[M^{-1}]_{ij} = \sum_\alpha \xi^\alpha_i 
\frac{1}{\lambda_\alpha + \delta_{\alpha 0}} \xi^\alpha_j 
= \sum_{\alpha\neq 0} \xi^\alpha_i 
\frac{1}{\lambda_\alpha} \xi^\alpha_j + \frac{1}{N}
\label{meq}
\end{equation}
and therefore 
\begin{equation}
b_i = \frac{1}{N (N - 1)} \sum_{k> l}\sum_j A_{ij} 
\vert M^{-1}_{ki} - M^{-1}_{kj} - M^{-1}_{li} + M^{-1}_{lj}\vert
+ \frac{1}{N}.
\label{betw2}
\end{equation}
Numerical results are obtained for various models using Eq.(\ref{betw2}).

Before concluding this subsection, we expand on the electrical
circuit analogy~\cite{newman}. For any graph, one can construct
a corresponding electrical circuit, where each edge is replaced by
a unit resistor. Instead of $r_i$ one has to obtain the voltage
$V_i$ at each node $i.$, The condition from Kirchoff's
laws is that $\sum_j (V_i - V_j) = 0$ for $i$ except nodes where
current enters or leaves the graph, where $j$ is summed over nearest
neighbors of $i.$ But this is identical to the condition on the
$r_i$'s in the random walk version of the problem, since the net
flow of walkers along the edge $(ij)$ is equal to $r_i - r_j,$ and
there is no net inflow in steady state at any node except the source
or the destination. Thus to obtain the random walk betweenness of
a node $i,$ we inject one unit of current at node $k$ in the circuit
and extract it at node $l,$ add the magnitudes of the currents
flowing through all the resistors connected to $i,$ and average
over $k$ and $l$ (with an extra factor of half and an additive
correction of $1/N$).

\subsection{Numerical results}
We first consider square and triangular lattice graphs with overall
shapes that are square and triangular respectively. Thus the square
lattice graphs have $L\times L$ nodes and the triangular lattice
graphs have $L(L+1)/2$ nodes, with various values of $L.$ The random
walk betweenness of all the nodes in a graph are sorted in decreasing
order to obtain the distribution of random betweenness for that
graph.

\begin{figure}
\begin{center}
\includegraphics[width=\columnwidth]{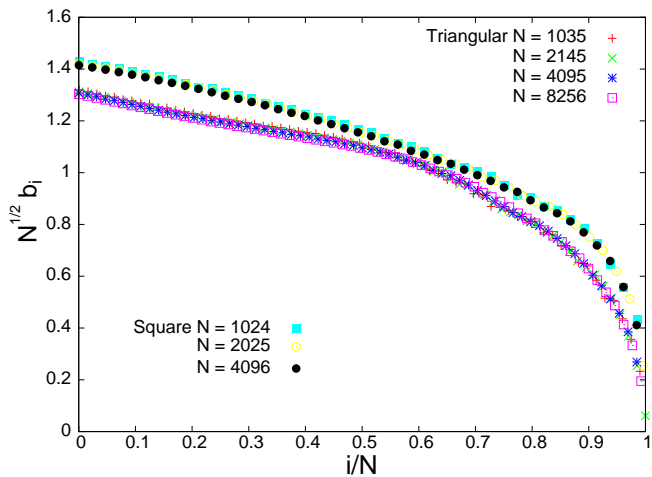}
\caption{Scaling collapse of the sorted random walk betweenness for
square and triangular lattice graphs. The form in Eq.(\ref{scollapseE})
is used. 
The vertical axis is multiplied by 1.05 for all the square 
lattice plots in order to separate the curves for the two lattices.}
\label{fig:scalingplot_lattices}
\end{center}
\end{figure}

One can view the graph as a discretization of a continuum square
or triangular surface.  As $N$ is increased, the mesh size is reduced
instead of the size of the surface being increased.  When one unit
of current is injected at one point and extracted at another, as 
$N$ is increased, the
current density approaches the form for a continuum surface, which 
we denote by ${\bf j}({\bf r}).$
In order to obtain the current distribution for the 
lattice graph from this continuum limit, we observe that the continuum
current flowing through a line segment of length $dl$ that is normal 
to the current flow ${\bf j}({\bf r})$ at a point ${\bf r}$ is 
$j({\bf r}) dl,$ and the number of resistors in the lattice graph 
that cut through the line segment $dl$ is $\sim \sqrt N dl.$ Since the 
current $j({\bf r}) dl$ is distributed over these resistors, the 
current in any one resistor is $\sim 1/\sqrt N f({\bf r}).$ If we sort
the nodes according to their betweenness, from the greatest to the least, 
in the $N\rightarrow \infty$ limit, the location ${\bf r}$ of the $i$'th 
node will only depend on the ratio $i/N.$ With the nodes thus reordered,
the sorted random walk betweenness should have the scaling form 
$b_i = \tilde b(i/N)/\sqrt N.$ This argument can be generalized for a 
$d$ dimensional lattice, yielding 
\begin{equation}
b_i = N^{1/d - 1} \tilde b (i/N).
\label{scollapseE}
\end{equation}
Numerical results for the square and triangular lattice are shown in 
Figure~\ref{fig:scalingplot_lattices}, and bear out this argument.

An alternative equivalent form of this result uses the fact that 
the average shortest path length between two randomly chosen nodes
in the graph is $l_N \sim N^{1/d},$ so that 
\begin{equation}
b_i = \frac{l_N}{N} \hat b(i/N).
\label{scollapse}
\end{equation}
This form can be tested for all graphs, including those for which there is 
no concept of discretization of a continuum manifold, and is found to work
there even though the argument given above does not apply.

We first consider Erdos Renyi~\cite{er} random graphs.  For any
pair of nodes $(ij)$ in the graph, the probability of their being
connected by an edge is $d_a/N.$ $d_a > 1$ is then the average nodal
degree in the graph.  There is also a dense regime for the Erdos
Renyi model, where $d_a$ is proportional to $\ln N,$ in contrast to 
the sparse regime where $d_a$ is $N$-independent. Because we
require that the graph should have a single component, only the
giant component of each graph is retained.

If the sorted random betweenness is plotted, there are large
fluctuations in the plot. This is due to the randomness in how the graphs are
constructed. In order to obtain a good scaling collapse,
eighty graph realizations are constructed for each $N,$ and the
betweenness of the $i$'th sorted node is averaged over these
realizations before a scaling collapse using Eq.(\ref{scollapse})
is attempted. (For large $N,$ the number of nodes in the giant
component of the graph tends to a definite fraction of $N$ in the
sparse regime, and $O(1)$ less than $N$ in the dense regime. Scaling
plots are therefore constructed with $N,$ rather than the actual
number of nodes in the giant component.)

\begin{figure}
\begin{center}
\includegraphics[width=\columnwidth]{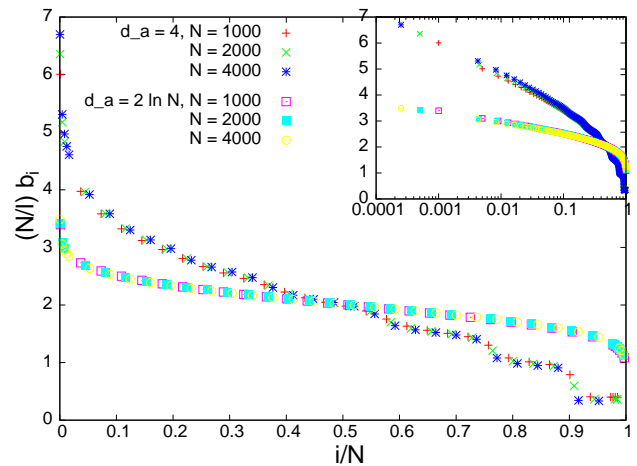}
\caption{Scaling collapse of the sorted random walk betweenness for
Erdos Renyi graphs, and two values of the average nodal degree
$d_a.$ Eq.(\ref{scollapse}) is used to construct the scaling plot. The 
inset shows the same distributions but with a log scale on the $x$ axis.}
\label{fig:scalingplot1_er}
\end{center}
\end{figure}

The average distance $l_N$ between nodes is found numerically for each
$N.$ Thus there are no adjustable parameters in the scaling plot.
The results for $d_a = 4$ in the sparse regime and $d_a = 2 \ln N$ in the 
dense regime  are shown in Figure~\ref{fig:scalingplot1_er}.
Despite
the fact that the justification given for Eq.(\ref{scollapse}) for square
and triangular lattices is not applicable here, a good scaling collapse is 
seen.

\begin{figure}
\begin{center}
\includegraphics[width=\columnwidth]{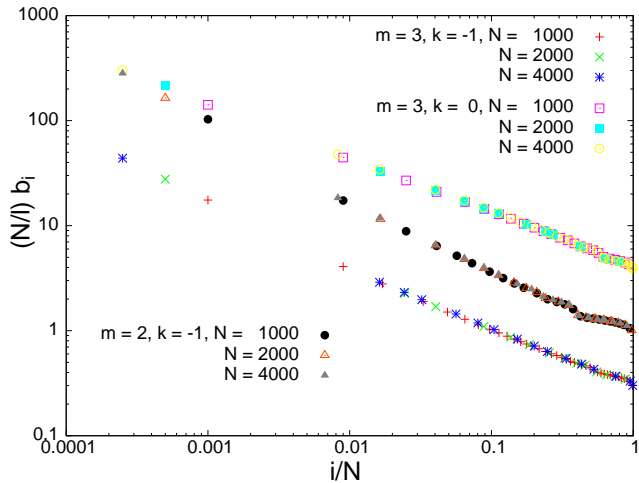}
\caption{Scaling collapse of the sorted random walk betweenness of
all the nodes in a scale-free graph. The three scaling plots
correspond to $(m,k) = (3,0), (2,-1)$ and $(3,-1).$ In order to separate
the plots, the betweenness in the first and third plots is 
divided and multiplied by 3.6 respectively.}
\label{fig:scalingplot_sf}
\end{center}
\end{figure}

Now we turn to preferential attachment graphs. Following the extension of
Ref.~\cite{redner} to the model of Ref.~\cite{barabasi}, nodes in
the network are created one by one, with each node born with $m$
edges that link it to preexisting nodes. The probability of linking
to a preexisting node of degree $m$ is proportional to $m+k,$ where
$k$ is a paramter of the model. As for Erdos Renyi graphs, eighty
realizations were constructed for each choice of graph parameters.
Again, although there is no obvious reason why the arguments leading 
to Eq.(\ref{scollapse}) should apply, the scaling collapse is 
very good for all the cases shown in Figure~\ref{fig:scalingplot_sf}.

The scaling of the sorted random betweenness has implications for
the node for which the random betweenness is maximum. From
Figure~\ref{fig:scalingplot_lattices}, the function $\hat b(i/N)$
is seen to have a well defined limit $\hat b(0)$ for the square and
triangular lattices. This implies that $b_{max}\sim l_N/N\sim 1/\sqrt
N.$ 
The situation
is more complicated for Erdos Renyi graphs; from the inset to
Figure~\ref{fig:scalingplot1_er}, the function $\hat b(i/N)$ is seen
to diverge logarithmically when its argument is small, at least in
the sparse regime. This implies that $b_{max}$ does not scale as
$\sim l_N/N.$ If one plots $b_{max}$ versus $N,$ one observes an 
apparent power law form with an exponent that only depends on 
whether one is in the dense (average degree $\sim\ln N$) or 
sparse (average degree independent of $N$) regime.
For preferential attachment graphs, Figure~\ref{fig:scalingplot_sf} shows that 
$\hat b(i/N)$ has a power law form when its argument is small, with
an exponent that changes when $m$ or $k$ changes. This means that 
--- ignoring the weak $N$-dependence in $l_N$ --- the
maximum random walk betweenness scales as a power of $N,$ with an
exponent that depends on $(m,k).$ If one plots $b_{max}$ versus 
$N,$ this is found to be the case with an exponent that only depends on 
$k/m.$

Finally, we consider the distribution of random walk betweenness
for hyperbolic grids, which are tilings of the Poincare disk~\cite{rno}.
For these graphs, it is not possible to obtain a scaling collapse
of the form given in Eq.(\ref{scollapse}): $b(i/N)$ is seen to be
$\sim 1/N$ when $i$ is comparable to $N,$ but is independent of $N$
when $i$ is small.  The maximum random betweenness $b_{max}$
approaches a limiting value as $N\rightarrow \infty$ for these
graphs. Despite the fact that these graphs are discretizations of
the Poincare disk, unlike the square and triangular lattices, the
underlying continuous surface has negative curvature.  As a result,
it is not possible to consider hyperbolic grids with different $N$'s
as being discretizations of the same continuous region but with
different mesh sizes, and the argument leading to Eq.(\ref{scollapse})
fails.
\begin{figure}
\begin{center}
\includegraphics[width=\columnwidth]{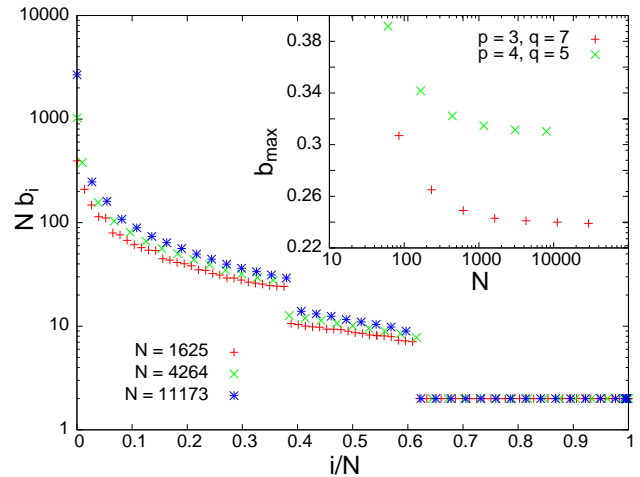}
\caption{Distribution of random walk betweenness for a hyperbolic
grid in which $q$ $p$ sided polygons meet at each node. The case
$(p,q) = (3,7)$ is shown. $N$ is varied by including all nodes
within a distance $r$ of a central node, and varying $r.$ The
function $N b(i/N)$ is plotted, and only the tail of the distribution
shows a scaling collapse. (Inset) Maximum random walk betweenness
for hyperbolic grids with $(p,q) = (3,7)$ and $(4,5)$  as a function
of $N,$ showing a non-zero $N\rightarrow\infty$ limit.}
\label{fig:betw_ht37}
\end{center}
\end{figure}

\section{Normalized random walk betweenness}
\label{sec:alternate}
The random walk betweenness defined in Ref.~\cite{newman} and studied
in the previous section ensures that if a random walk goes back and
forth between two nodes, this does not increase their betweenness.
However, if the walk goes round a loop repeatedly, the betweenness of all
the nodes in the loop is increased by unity for each round-trip. If one is interested in finding
whether a node lies in the path of many random walks, it would appear to be better
to define the betweenness so that each random walk contributes
exactly once to the nodes it passes through, regardless of how many
times it passes through them. This quantity can be studied using 
the technique of the previous section, with a slight modification. To avoid 
confusion, we call this quantity the normalized random walk betweenness. 

As before, we consider random walkers from a source node $k$ to a
destination node $l,$ and try to obtain the normalized random betweenness of
some query node $m.$ At every time step, one walker is injected into the network at
the node $k.$ If the walker reaches the destination
node $l$ {\it or\/} the query node $m,$ 
it disappears from the network at the next time step. Clearly, any
random walk that reaches $m$ before it reaches $l$ contributes
once to the extinction at the $m$'th node, even if it would
have gone through $m$ multiple times before reaching $l$ if it had been 
allowed to continue. Any node that reaches $l$ without having gone through
$m$ does not contribute to the extinction at $m.$ Thus the rate at which walkers
are destroyed at the node $m$ is the probability that a random walker from $k$ to 
$l$ will pass at least once through $m.$ 

In steady
state, in place of Eq.(\ref{ee3}), we have
\begin{equation}
p^{kl}_i= \delta_{ik} - (1 - \mu^{kl}_m)\delta_{il} -\mu^{kl}_m\delta_{im} 
   + \sum_j A_{ij} \frac{p^{kl}_j}{d_j}
\label{een3}
\end{equation}
where $\mu^{kl}_m$ is the rate at which random walkers from $k$ to
$l$ reach $m$ (for the first time), which has to be determined
self-consistently. Eq.(\ref{een3}) comes with the boundary conditions
$p^{kl}_l = p^{kl}_m = 0.$ 

With $p^{kl}_i = d_i r^{kl}_i,$ we obtain
the solution
\begin{equation}
r^{kl}_j = \sum_{\alpha = 1}^{N-1} 
\frac{\xi^\alpha_k - (1 - \mu^{kl}_m)\xi^\alpha_l - \mu^{kl}_m\xi^\alpha_m}
{\lambda_\alpha}\xi^\alpha_j + c^{kl} \xi^0_j.
\end{equation}
Applying the condition $r^{kl}_m - r^{kl}_l = 0,$ we obtain 
\begin{equation}
\sum_{\alpha = 1}^{N-1} 
\frac{\xi^\alpha_k - (1 - \mu^{kl}_m)\xi^\alpha_l - \mu^{kl}_m\xi^\alpha_m}
{\lambda_\alpha} (\xi^\alpha_m - \xi^\alpha_l) = 0
\end{equation} 
which fixes $\mu^{kl}_m:$
\begin{equation}
\sum_{\alpha = 1}^{N-1} 
\frac{\xi^\alpha_k - \xi^\alpha_l}{\lambda_\alpha}
(\xi^\alpha_m - \xi^\alpha_l) = 
\mu^{kl}_m\sum_{\alpha = 1}^{N-1} 
\frac{(\xi^\alpha_m - \xi^\alpha_l)^2}{\lambda_\alpha}.
\end{equation}
Note that $\mu^{kl}_k = 1,$ as it should be. The expression for
$\mu^{kl}_l$ is indeterminate; we fix it to be 1.

In order to obtain the normalized random walk betweenness $n_m$ of
the node $m,$ we average $\mu^{kl}_m$ over all $l$ and $k\neq l,$
with $m$ fixed.  Thus
\begin{equation}
n_m = [\sum_{l\neq m} \sum_{k\neq l} \mu^{kl}_m + (N-1)]/(N(N-1)).
\end{equation}
In the numerator, the sum over $k$ can be made unrestricted, because
$\mu^{ll}_m = 0$ if $l\neq m.$ Since $\sum_k \xi^\alpha_k = 0$ for
$\alpha\neq 0,$ we have
\begin{equation}
\sum_k \mu^{kl}_m = N  \bigg[\sum_{\alpha\neq 0} 
\frac{(\xi^\alpha_l - \xi^\alpha_m)\xi^\alpha_l}{\lambda_\alpha}
\bigg] \bigg[\sum_{\alpha\neq 0} 
\frac{(\xi^\alpha_l - \xi^\alpha_m)^2}
{\lambda_\alpha}\bigg]^{-1}.
\end{equation}
Therefore 
\begin{eqnarray}
n_m &=& \frac{1}{N-1} \sum_{l\neq m}  \bigg[\sum_{\alpha\neq 0} 
\frac{(\xi^\alpha_l - \xi^\alpha_m)\xi^\alpha_l}{\lambda_\alpha}
\bigg] \bigg[\sum_{\alpha\neq 0} 
\frac{(\xi^\alpha_l - \xi^\alpha_m)^2}{\lambda_\alpha}\bigg]^{-1}\nonumber\\
&+& \frac{1}{N}.
\end{eqnarray}
Using Eq.(\ref{meq}) we have 
\begin{equation}
M^{-1}_{ll} - M^{-1}_{ml} = \sum_{\alpha\neq 0} 
\frac{(\xi^\alpha_l - \xi^\alpha_m) \xi^\alpha_l}{\lambda_\alpha}
\end{equation}
from which 
\begin{equation}
n_i =  
\frac{1}{N} + 
\frac{1}{N-1} \sum_{l\neq i} 
\frac{M^{-1}_{ll} - M^{-1}_{li}}
{M^{-1}_{ll} + M^{-1}_{ii} - 2 M^{-1}_{li}}
\label{norm_m}
\end{equation}
where we have replaced the subscript $m$ with $i$ to match the
expression for the random walk betweenness $b_i$ in
Section~\ref{sec:rwbetw}.

\subsection{Numerical results}
\begin{figure}
\begin{center}
\includegraphics[width=\columnwidth]{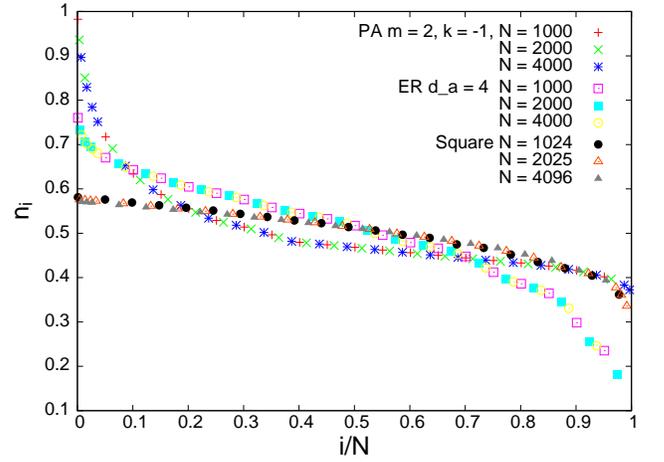}
\caption{Normalized random walk betweenness for various graph models.
The nodes are sorted according to decreasing betweenness. Results
for preferential attachment graphs with $m = 2, k = -1,$ Erdos Renyi graphs with
$d_a = 4,$ and square lattice graphs of various sizes are shown.
Except for the square lattices, each data point represents the
average of eighty random graphs.}
\label{fig:sclplt}
\end{center}
\end{figure}
\begin{figure}
\begin{center}
\includegraphics[width=\columnwidth]{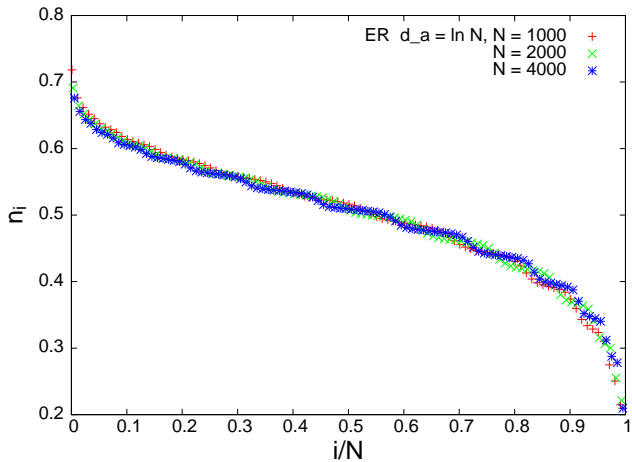}
\caption{Normalized random walk betweenness for the Erdos Renyi
model in the dense regime with the graph size $N$ varied: the average
nodal degree is $d_a = \ln N.$ The betweenness of all the nodes is
sorted, and an average over eighty random graphs is taken. The
results for different $N$ are approximately on the same curve, but
`ripples' are seen at different places. }
\label{fig:sclplt_er}
\end{center}
\end{figure}

Eq.(\ref{norm_m}) was used to numerically evaluate the normalized
random walk betweenness $n_i$ for various models. As in
Section~\ref{sec:rwbetw}, the $n_i$ values for all the nodes in a
graph were sorted. For the random graph models, the sorted list was
averaged over eighty realizations of the random graph.
Figure~\ref{fig:sclplt} shows the normalized random walk betweenness
for preferential attachment networks, the Erdos Renyi model in the sparse regime,
and lattice graphs. For all these cases, $n_i$ is only a function
of $i/N.$ For clarity, the figure only shows one example of each
class of graph models, but a similar data collapse (to different
curves) is seen if the model parameters are varied. For the 
Erdos Renyi model in the dense regime, the scaling collapse is 
imperfect, as shown in 
Figure~\ref{fig:sclplt_er}. 

\begin{figure}
\begin{center}
\includegraphics[width=\columnwidth]{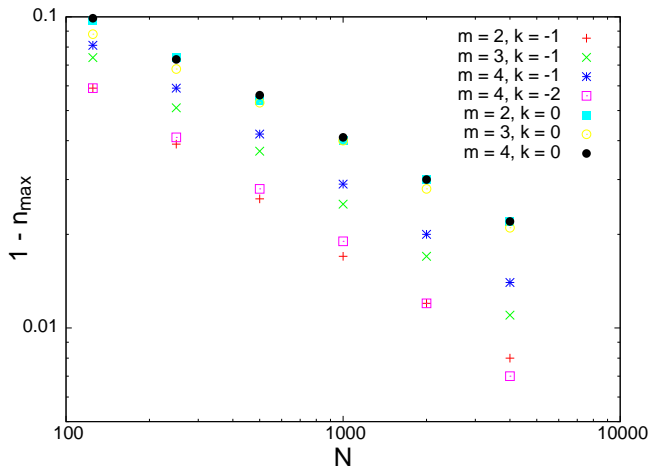}
\caption{Deviation from unity  of the maximum normalized random
walk betweenness, $n_{max},$ for preferential attachment graphs with various
parameters. The plots show that $1 - n_{max}$ decreases as the size
of the graph $N$ is increased as $\sim 1/N^c,$ with the exponent
$c$ depending on the ration of model parameters, $k/m.$}
\label{fig:nmax}
\end{center}
\end{figure}

In Figure~\ref{fig:sclplt}, we see that the maximum normalized
betweenness is close to 1 for the preferential attachment graphs. In fact, as
seen in Figure~\ref{fig:nmax}, $n_{max}\rightarrow 1$ as
$N\rightarrow\infty$ for these models. This may seem surprising:
$n_{max} = 1$ implies that {\it all\/} the random walks pass through
some node. When shortest path routing is used instead of random
walks, this is impossible, and one might expect even less `focusing'
on any node with random walk routing.

To understand how this could happen, we consider a simpler graph:
a regular tree that descends $h$ levels from a root node.  Except
at the lowest level, each node is connected to $q$ nodes at the
next level. With randomly chosen source and destination nodes $k$
and $l$, we calculate the probability that a random walk from $k$
to $l$ will pass through the root node. Let $m$ be the common
ancestor of $k$ and $l.$ When the random walk first reaches $m,$
we have to calculate the probability that, thereafter, it reaches
the root node before it reaches its destination. To obtain this
probability, all we really need is the straight line path from the
root node through $m$ to $l;$ all other parts of the tree are
detours, which affect the time it takes to reach the root node or
its destination, but not the probability. For a random walk on this
straight line, it is easy to calculate that the desired probability
to reach the root node first is is $d(m,l)/[d(m,l) + d(0,m)] = 1 -
d(0,m)/d(0,l),$ where $d(i,j)$ is the distance between nodes $i$
and $j.$ The probability distribution for $d(0, l)$ is $q^{d(0,
l)}/[q^h (1 - 1/q)],$ so that $h - d(0, l)$ is typically $O(1).$
Then for large $h,$ the probability distribution for $d(0, m)$ is
$q^{-d(0, m)}(1 - 1/q),$ so that $d(0, m)$ is also typically $O(1).$
Then $\langle d(0, m)/d(0, l)\rangle \sim 1/h\sim 1/\ln N.$ Therefore
the probability that a random walk between randomly chosen source
and destination nodes passes through the root node is $1 - O(1/\ln
N).$ 

From Figure~\ref{fig:nmax} we see that the root node is even
more central for random walks on preferential attachment graphs than on trees,
with $1 - n_{max}\sim 1/N^c.$ Also surprising is the observation that 
$n_{max}(N\rightarrow\infty)$ is not unity for Erdos-Renyi graphs 
even though they are locally tree-like.

\section{Conclusions} 
To summarize, we have considered two definitions of the random walk
betweenness of the nodes in a graph: an unnormalized version which
can be represented in terms of currents flowing in an electrical
circuit, and a normalized version which does not measure the number
of passes for a random walk through a node but instead uses a binary
measure that denotes passage.  For both cases, we obtain a parameter
free scaling collapse of the distribution of node betweenness as a
function of the size $N$ of the graph, for several graph models ---
with the exception of hyperbolic grids.  The scaling function is
singular for scale-free graphs, resulting in the maximum unnormalized
betweenness being of the form $\sim 1/N^a$ with nontrivial values
for $a.$ Although the scaling collapse can be understood for lattice
graphs through a continuum limit, it is not clear why it
works for random graph models.

\section{Acknowledgements}
\begin{acknowledgments}
This work was supported by grants FA9550-11-1-0278 and 
60NANB10D128 from AFOSR and NIST, respectively.
\end{acknowledgments}

\end{document}